\documentclass[conference]{IEEEtran}
\IEEEoverridecommandlockouts
\usepackage{multirow}

\usepackage{cite}
\usepackage{amsmath,amssymb,amsfonts}
\usepackage{algorithmic}
\usepackage{graphicx}
\usepackage{url}
\usepackage{textcomp}
\usepackage{xcolor}
\usepackage{hyperref}
\usepackage{fancyhdr}
\def\BibTeX{{\rm B\kern-.05em{\sc i\kern-.025em b}\kern-.08em
    T\kern-.1667em\lower.7ex\hbox{E}\kern-.125emX}}
\begin{document}

\title{Bottleneck Transformer-Based Approach for Improved Automatic STOI Score Prediction\\
}

\author{\IEEEauthorblockN{1\textsuperscript{st} Amartyaveer}
\IEEEauthorblockA{\textit{Spire Lab, Dept of Electrical Engg} \\
\textit{Indian Institue of Science (IISc)}\\
Bengaluru, India \\
Amartyaveer72@gmail.com}
\and
\IEEEauthorblockN{2\textsuperscript{nd} Murali Kadambi}
\IEEEauthorblockA{\textit{Spire Lab, Dept of Electrical Engg} \\
\textit{Indian Institue of Science (IISc)}\\
Bengaluru, India \\
mkkadambi@gmail.com}
\and

\IEEEauthorblockN{3\textsuperscript{rd} Chandra Mohan Sharma}
\IEEEauthorblockA{\textit{Center for Artificial Intelligence } \\
\textit{and Robotics (CAIR), DRDO} \\
India \\
chandramohan.cair@gov.in}
\and
\IEEEauthorblockN{4\textsuperscript{th} Anupam Mandal}
\IEEEauthorblockA{\textit{Center for Artificial Intelligence} \\ 
\textit{and Robotics (CAIR), DRDO} \\
India \\
amandal.cair@gov.in}
\and
\IEEEauthorblockN{5\textsuperscript{th} {Prasanta Kumar Ghosh}}
\IEEEauthorblockA{\textit{Spire Lab, Dept of Electrical Engg} \\
\textit{Indian Institue of Science (IISc)}\\
Bengaluru, India \\
prasantg@iisc.ac.in}

}

\maketitle

\begin{abstract}

In this study, we have presented a novel approach to predict the Short-Time Objective Intelligibility (STOI) metric using a bottleneck transformer architecture. Traditional methods for calculating STOI typically require clean reference speech. This limits their applicability in the real world. To address this, numerous deep learning-based non-intrusive speech quality assessment models have garnered significant interest. Many studies have achieved commendable performance, but there is room for further improvement.
    
We propose the use of bottleneck transformer, convolution blocks for learning frame-level features, and a multi-head self-attention (MHSA) layer to aggregate the information. These components enable the transformer to focus on key aspects of the input data. The proposed model, in spite of having relatively fewer parameters, has shown higher correlation and lower mean squared error in STOI prediction for both seen and unseen scenarios compared to the state-of-the-art model using self-supervised learning (SSL) and spectral features as input.
\end{abstract}

\fancypagestyle{plain}{
  \fancyhf{}
  \fancyfoot[C]{\footnotesize © 2025 IEEE. Personal use of this material is permitted. Permission from IEEE must be obtained for all other uses, in any current or future media, including reprinting/republishing this material for advertising or promotional purposes, creating new collective works, for resale or redistribution to servers or lists, or reuse of any copyrighted component of this work in other works.}
  \renewcommand{\headrulewidth}{0pt}
  \renewcommand{\footrulewidth}{0pt}
}
\thispagestyle{plain}


\section{Introduction}
Speech quality assessment (SQA) \cite{kondo2025rethinking} refers to the process of evaluating various attributes of speech signals, such as quality and intelligibility. SQA metrics/scores are indicators that quantitatively measure the specific attributes of speech signals. SQA is mainly divided into two categories, one that requires human involvement called Subjective Assessment, and one that does not require human listeners called Objective Assessment. Objective assessment is further divided into two subcategories called intrusive assessment and non-intrusive assessment. The former requires a clean reference signal to calculate the assessment score, while the latter does not require the clean reference signal. For most of the cases where we have a large amount of real-world noisy recordings, the clean reference signal is unavailable, and therefore, subjective or intrusive assessment methods are not feasible. To overcome this limitation, several approaches have been proposed to estimate speech intelligibility instead of human listening tests and intrusive assessment.

Fu et al. in \cite{fu2018quality} proposed Quality-Net to predict speech quality. They used the spectrogram as an input to bidirectional long-short-term memory (BiLSTM) modules. They proposed to estimate the Perceptual evaluation of speech quality (PESQ) score \cite{941023} at an utterance level by incorporating the weighted sum of both the utterance level and frame level estimations using the mean squared error (MSE) as the objective function. Zezario et al. in \cite{zezario2020stoi} introduced STOI-Net which also used the spectrogram as an input. STOI-Net architecture is a combination of convolutional neural networks (CNN) and BiLSTM modules with multiplicative attention mechanism (CNN-BiLSTM-ATTN). An objective function identical to that employed by Quality-Net \cite{fu2018quality} was used, whose predictions showed a higher correlation with ground-truth STOI scores.

More research works were published with multi-task setup where scores like speech transmission index (STI), STOI\cite{taal2010short}, PESQ and human-subjective ratings from human listening tests were predicted. MOSA-Net\cite{zezario2022deep} used cross-domain features (spectral and temporal features) and latent representations from SSL model, namely HuBERT \cite{hsu2021hubert} to predict objective quality and intelligibility scores simultaneously. MOSA-Net can quite accurately predict PESQ and STOI scores, with Linear Correlation Coefficient (LCC) up to $98.8\%$ for PESQ and $97.7\%$ for STOI. Later an improved version of MOSA-Net called MTI-Net\cite{zezario122022mti} was developed to simultaneously predict Subjective Intelligibility (SI), STOI and Word Error Rate (WER) scores. 

Several works have been reported for predicting Mean Opinion Score (MOS). Recent efforts include MOSNet\cite{lo2019mosnet}, a CNN-BiLSTM-based model designed to estimate MOS score. MBNET\cite{leng_2021_mbnet} uses two subnets, MeanNet, to predict the mean score and BiasNet, to predict the bias from the mean score. Input to MBNET are the audio spectrum and judge id; clipped MSE is part of the loss function for training the two subnets together. The mean and bias scores are added to arrive at the prediction of the MOS score. QUAL-Net\cite{lin2024non} utilizes the same architecture and features as MTI-Net\cite{zezario122022mti} for MOS prediction but uses a simpler CNN architecture for feature extraction.

More work has been done on clinical speech assessment. DNN-based models are utilized in hearing aids (HA) to predict evaluation metrics such as the Hearing Aid Speech Quality Index (HASQI) \cite{kates2010hearing} and the Hearing Aid Speech Perception Index (HASPI)\cite{kates2014hearing}.
MBI-Net \cite{edozezario22_interspeech} resembles MTI-Net \cite{zezario122022mti} and takes spectral feature along with a hearing loss pattern as inputs for the network. It has two branches that uses different input channels, fed into a feature extractor to extract spectral features, learnable filter banks (LFB), and SSL features and estimates subjective intelligibility scores. MBI-Net+\cite{zezario24_interspeech}, an enhanced version, incorporates HASPI into its objective function to enhance the intelligibility prediction score. It uses Whisper model embeddings and speech metadata as inputs and utilizes a classifier to identify speech signals enhanced by various methods.

In this study, we propose a model for STOI prediction. The proposed model comprises of a convolution block (conv block), a bottleneck transformer \cite{9577771}, and dense layers. The conv block is used for extracting and refining the input features. The bottleneck transformer helps to capture short- and long-term contexts while removing redundant information. The dense layer is used for the prediction of the STOI scores. Based on our research among recent model architectures, STOI-Net \cite{zezario2020stoi} was one of the only models that predicted only one metric as the output i.e. STOI. Therefore, we used the STOI-Net model as our baseline model. Experimental results show that predicted scores of the proposed model have higher correlation with the ground-truth STOI Scores when tested in both Seen (as explained in Section V) and Unseen conditions (test speakers and utterances are not involved in the training) than the baseline model. 


\section{Dataset}
Due to the unavailability of datasets containing STOI scores, we developed our own dataset, where noise has been added to clean recordings to create synthetic noisy audio samples.
We selected the Indic TIMIT \cite{yarra2019indic},  Librispeech\cite{panayotov2015librispeech}, RESPIN \cite{singh2023model} and Bhashini\footnote{https://bhashini.gov.in} Hindi datasets to obtain speech recordings. The levels of Signal-to-Noise Ratio (SNR) in the audio clips from the datasets were analyzed using waveform amplitude distribution analysis (WADA)-SNR\cite{kim08e_interspeech} metric. Files with WADA-SNR estimations exceeding 80dB were classified as clean recordings and used to build the noisy dataset. Various types of noises were added to the clean recordings to form a comprehensive dataset of noisy recordings. The following is the list of distortions used:
\begin{enumerate}
\item \textbf{Mobile/Telephone channel noise}: For simulating the mobile/telephone channel, we used SoX which simulates the characteristics of the Global System for Mobile Communication (GSM) format with SoX.
\item \textbf{Reverberation}: Reverberation was induced in speech signals by convolving the signals, with Room Impulse Responses (RIR) from the ACE Corpus \cite{7486010}. The available data in this corpus contain real recordings of RIRs with their corresponding T-60 values from 7 different rooms with varying positions of microphones.
\item \textbf{Radio channel noise}: To simulate radio channel noise, The signal was passed through a band pass filter with one cut-off frequency between 50-1000Hz and second cut-off at 2600Hz. White noise of 30-40 dB SNR was added after this.  
\item \textbf{Trans-coding (mp3, ogg, flac, aiff, wav)} : In transcoding we converted the raw audios in different audio codec formats. Each audio codec has its own way of compressing the audio, which may affect the intelligibility of the audio. After compression, we decompress the audio back to wav format so that the final wav format contains the effect of the codecs used for compression.
\item \textbf{Variable length clipping}: Clipping affects intelligibility a lot \cite{clipping_9281027}. We used a moving window, within which we randomly selected negative and positive thresholds and clipped the samples having values higher or lower than the threshold.
\item \textbf {Additive Noise}: A randomly selected noise from the MUSAN \cite{musan2015} dataset like, white noise, pink noise, machine gun noise, etc. is added to a clean recording at a random SNR in the range of $0-20$ dB. 
\end{enumerate} 

We added these aforementioned noises and distortions in three combinations. The resulting noisy files have distortions of either a single type of noise, a combination of two noises, or a combination of three noises. To compute the ground-truth STOI score between the noisy signal and the clean reference, we utilized the STOI metric from TorchMetrics Audio\footnote{\href{https://torchmetrics.readthedocs.io/en/v0.9.0/audio/short_time_objective_intelligibility.html}{TorchMetrics documentation on STOI}}.
 
We used a 12-hour Indic TIMIT subset for training, validation, and testing. This subset was divided into six parts of 2 hours each. One of the 2-hour part was held out as the Seen test set, while the other five parts were used for training and validating five models (refer to Section V). The Seen test set consisted of data in which speakers, utterances, and noise types from the remaining five parts appear in various combinations in the test set. The same combination of the three was ensured not to be present in the test set, thus preventing data leakage. Additionally, four unseen test sets were created by using 2 hours each from Librispeech, RESPIN (Bhojpuri and Bengali subset) and Bhashini (Hindi language) datasets and adding the aforementioned distortions to them. These Unseen test sets consisted of speakers and utterances different from those used in the training data. However, here, the noise types still overlapped with those used in the training data.



\section{Existing Methods for STOI Prediction}
ML-based SQA models are increasingly used in speech processing, including building SQA systems. This section reviews deep learning techniques for predicting speech metrics. Zezario et al. \cite{zezario122022mti} proposed MTI-Net, a network that predicts STOI, WER, and intelligibility in separate branches. It uses Short-Time Fourier Transform (STFT) features, LFB, and embeddings from an SSL model like HuBERT. The network comprises of convolution layers, BiLSTM layers, and linear layers, with predictions occurring in individual branches. The loss function combines utterance level and frame-level scores for each metric it predicts. The results showed that a model that uses all three feature types to predict all metrics was the best model. They also reported that replacing the pre-trained SSL model with a fine-tuned one further improved performance.

Zezario et al. \cite{zezario2022deep} introduced MOSA-Net, a network that used the embeddings as STFT, LFB, and SSL features with an architecture similar to MTI-Net \cite{zezario122022mti}. MOSA-Net predicted PESQ, STOI, and SDI (Speech Distortion Index) scores. Using the Wall Street Journal (WSJ) dataset, it was shown that MOSA-Net performed well in predicting PESQ scores with a BiLSTM + CNN + Attention configuration when noise types were seen during training. This setup also excelled at predicting STOI for unseen noise types. Additionally, they presented QIA-SE, a network that enhances noisy speech using the output of the final linear layer of MOSA-Net as an input.

Zezario et al. \cite{zezario2020stoi} introduced a network called STOI-Net that used the STFT spectrum as input feature for the model. The model was built with convolution,  BiLSTM and attention layers. The model showed a good correlation with actual STOI scores while performing non-intrusive prediction on noisy WSJ dataset.

Drgas et al. \cite{DRGAS2024103068} used Generalized Enhanced STOI containing an LFB that learns to process temporal envelopes. The Temporal Attention Block then weighs the contribution of the input speech in different time segments. The work in \cite{10447597} used features from the whisper ASR model's decoder layers for the prediction of intelligibility in hearing aids.  The work in \cite{10870355} predicted the intelligibility of speech due to hearing loss and hearing aids. A pre-trained WavLM model\cite{wavLM_paper} was used to extract acoustic features for each audio segment. These were then averaged across the time dimension to produce a representative speech feature vector.
Tamm et al. \cite{tamm22_interspeech} used acoustic features of a pre-trained wav2vec-based XLS-R model as input to a machine learning network to predict the MOS score. The network consisted of BiLSTM, attention, pooling and linear layers. A comparison in \cite{tamm22_interspeech} showed that the acoustic features of the XLS-R model using BiLSTM and Attention layers gave the best performance and also performed better than the two baseline models considered. Cooper et al. \cite{9746395} experimented with several fairseq models by fine-tuning and zero-shot training and found that the small and large variants of the wav2vec 2.0 model \cite{baevski2020wav2vec} gave the best and most consistent results.  

\section{Proposed Method}
\begin{figure}[!t]
    \centering
    \includegraphics[width=0.9\columnwidth]{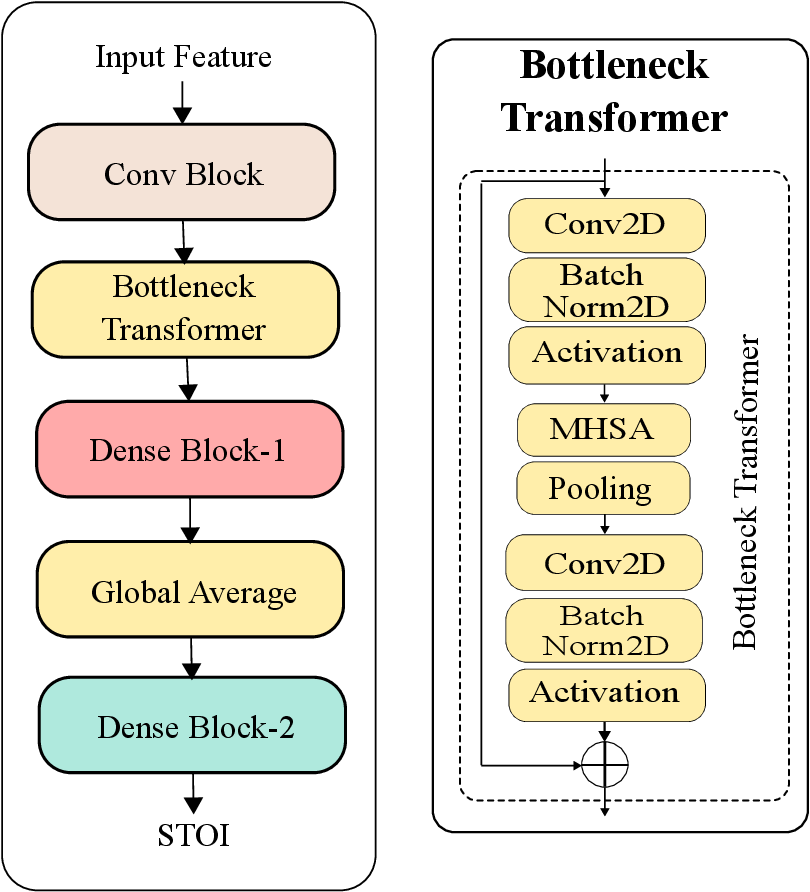}
    \caption{Proposed Architecture: Left- Architecture of the full model. Right- Architecture of the Bottleneck Transformer}
    \label{fig:model}
\end{figure}

The overall architecture of our proposed method is shown in \autoref{fig:model}. The input feature is processed through five main blocks, namely, Conv Block, Bottleneck Transformer, Dense Block-1, Global Average, and Dense Block-2. Three types of input features were experimented with for the model. 
\begin{enumerate}
    \item Latent Feature Vector from the projection layer of SSL models, like wav2vec 2.0 \cite{baevski2020wav2vec} small and HuBERT\cite{hsu2021hubert} base.
    \item Spectral features similar to the feature extraction method from STOI-Net\cite{zezario2020stoi} where the utterance was converted into a 257-dimensional spectrogram using a 512-point STFT with a 32 ms Hamming window and a 16 ms hop length. These shall be referred to as PS-I.
    \item Post-processing PS-I using a set of convolution layers to extract features, leveraged from \cite{zezario2020stoi} to be called as PS-II; and a set of convolution layers leveraged from \cite{lin2024non} to be referred to as PS-III. 
\end{enumerate}
  These features were learnable and directly fed into the convolution block (Conv Block) of the proposed model. It should be noted that since baseline is STOI-Net model, passing PS-I features into the baseline would automatically result in the model getting PS-II as input. Hence, PS-I shall not be listed as one of the input features to the baseline model.

\textbf{Conv Block}:
The Conv block consists of a set of two 1D convolutional layers along with a 1D Batch Normalization layer for regularization and a Gaussian Error Linear Unit (GELU)  \cite{hendrycks2023gaussianerrorlinearunits} as the activation. It takes input features, reduces the input dimension, and helps in extracting and refining the features. The extracted features are then provided as input to the Bottleneck Transformer.

\textbf{Bottleneck Transformer}:
For predicting STOI, we believe the model needs to capture both local and global information present even in the presence of non-stationary noise or distortions in the recording. For this purpose, we use a bottleneck transformer\cite{9577771}. It includes convolution layers, MHSA layers, Batch Normalization, and Pooling layers. Using both convolution and attention layers, it captures a richer and more robust representation from its input. Convolution layers reduce dimensionality and capture the local context, while self-attention layers aggregate the information learned by convolution layers and learn the global context. Additionally, another convolution layer is used for up-sampling. These layers filter out redundant information and focus on relevant data. A residual connection from input to output aids in gradient propagation during training. The bottleneck transformer’s output becomes input to Dense Block-1.

\textbf{Dense Blocks}
Dense blocks comprise linear layers and nonlinear activation layers. Dense Block-1 further refines the features extracted by the bottleneck transformer. Global pooling is then applied to reduce the temporal dimensions by averaging across the time dimension. This input is fed into Dense Block-2, which predicts the STOI score.

The objective function used for training was the MSE between the true and predicted utterance level STOI scores. The model effectively captures information at both the frame and the utterance levels without the need for frame-level scores for training.

\section{Experiments and Results}
 Five models of the proposed and baseline architectures were trained. Data from the five parts of the Indic TIMIT subset were rotated in a round-robin fashion, such that four parts were the training data and one part was the validation data, in each round resulting in each of the five models. 


\subsection{Experimental Setup}

\begin{table}[]
\centering
\begin{tabular}{|c|c|c|}
\hline
{\textbf{Features}} & {\textbf{Proposed Model}} & {\textbf{STOI-Net}} \\
                                                                     \hline
\multirow{2}{*}{\textbf{wav2vec 2.0}} & \multirow{2}{*}{727,233}                 & \multirow{2}{*}{967,938}           \\
                                   &                                          &                                    \\ \hline
\multirow{2}{*}{\textbf{HuBERT}}   & \multirow{2}{*}{923,841}                 & \multirow{2}{*}{1,230,082}         \\
                                   &                                          &                                    \\ \hline
\multirow{2}{*}{\textbf{PS-I}}     & \multirow{2}{*}{334,785}                 & \multirow{2}{*}{N/A}               \\
                                   &                                          &                                    \\ \hline
\multirow{2}{*}{\textbf{PS-II}}    & \multirow{2}{*}{1,019,937}               & \multirow{2}{*}{1,195,106}         \\
                                   &                                          &                                    \\ \hline
\multirow{2}{*}{\textbf{PS-III}}   & \multirow{2}{*}{669,921}                 & \multirow{2}{*}{845,538}           \\
                                   &                                          &                                    \\ \hline
\end{tabular}
\vspace{0.4em} 
\caption{Parameter Count of Models for different feature types}
\label{tab:param_count}
\end{table}

\begin{table}[]
\resizebox{\columnwidth}{!}{%
\setlength{\tabcolsep}{4pt}
\begin{tabular}{|c|c|c|c|c|}
\hline
\textbf{Model} & \textbf{Features} & \textbf{LCC ↑} & \textbf{SRCC ↑} & \textbf{MSE ↓} \\ \hline

\multirow{8}{*}{\textbf{STOI-Net}} 
& \multirow{2}{*}{\textbf{wav2vec 2.0}} & \multirow{2}{*}{$93.25 \pm 0.99$} & \multirow{2}{*}{$94.74 \pm 1.16$} & \multirow{2}{*}{$0.0075 \pm 0.0011$} \\
& & & & \\ \cline{2-5}
& \multirow{2}{*}{\textbf{HuBERT}}   & \multirow{2}{*}{$90.87 \pm 1.42$} & \multirow{2}{*}{$93.71 \pm 0.71$} & \multirow{2}{*}{$0.098 \pm 0.0016$} \\
& & & & \\ \cline{2-5}
& \multirow{2}{*}{\textbf{PS-II}}    & \multirow{2}{*}{$93.39 \pm 0.52$} & \multirow{2}{*}{$94.54 \pm 0.42$} & \multirow{2}{*}{$0.0071 \pm 0.0005$} \\
& & & & \\ \cline{2-5}
& \multirow{2}{*}{\textbf{PS-III}}   & \multirow{2}{*}{$78.78 \pm 7.46$} & \multirow{2}{*}{$79.36 \pm 7.27$} & \multirow{2}{*}{$0.0252 \pm 0.0083$} \\
& & & & \\ \hline

\multirow{10}{*}{\textbf{Proposed}}
& \multirow{2}{*}{\textbf{wav2vec 2.0}} & \multirow{2}{*}{$94.38 \pm 0.30$} & \multirow{2}{*}{$\mathbf{95.88 \pm 0.29}$} & \multirow{2}{*}{$0.0061 \pm 0.0003$} \\
& & & & \\ \cline{2-5}
& \multirow{2}{*}{\textbf{HuBERT}}   & \multirow{2}{*}{$\mathbf{94.63 \pm 0.52}$} & \multirow{2}{*}{$95.67 \pm 0.40$} & \multirow{2}{*}{$\mathbf{0.0059 \pm 0.0007}$} \\
& & & & \\ \cline{2-5}
& \multirow{2}{*}{\textbf{PS-I}}     & \multirow{2}{*}{$90.85 \pm 0.35$} & \multirow{2}{*}{$92.91 \pm 0.54$} & \multirow{2}{*}{$0.097 \pm 0.0003$} \\
& & & & \\ \cline{2-5}
& \multirow{2}{*}{\textbf{PS-II}}    & \multirow{2}{*}{$90.39 \pm 1.40$} & \multirow{2}{*}{$92.64 \pm 1.48$} & \multirow{2}{*}{$0.0101 \pm 0.0013$} \\
& & & & \\ \cline{2-5}
& \multirow{2}{*}{\textbf{PS-III}}   & \multirow{2}{*}{$93.50 \pm 0.35$} & \multirow{2}{*}{$94.96 \pm 0.25$} & \multirow{2}{*}{$0.0071 \pm 0.0004$} \\
& & & & \\ \hline

\end{tabular}}
\vspace{0.1em}
\caption{Baseline and proposed model performance on the seen test set}
\label{tab:seen}
\end{table}

\begin{figure*}[!ht]
    \centering
    \includegraphics[width=1.9\columnwidth]{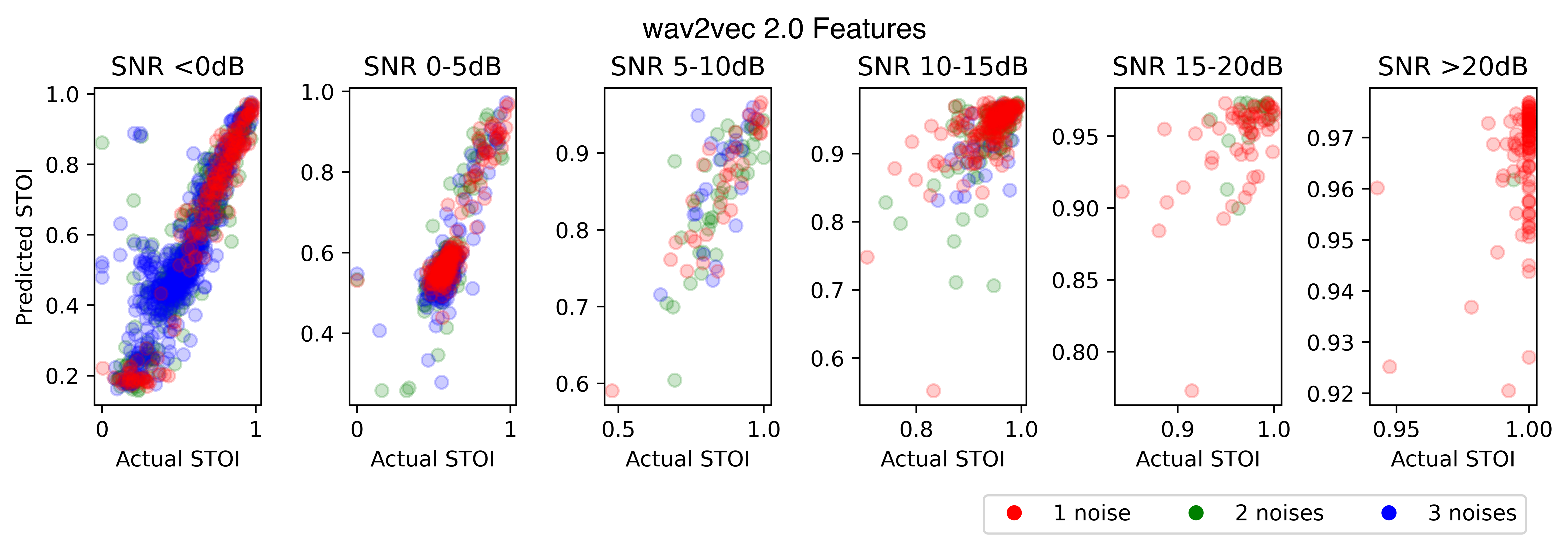}
    \caption{Predicted STOI vs. Actual STOI for various SNR bins when using wav2vec 2.0 Features for seen test set using proposed model}
    \label{fig:predicted_vs_actual_stoi_across_snr_bins_W2V2_Features}
\end{figure*}

\begin{table*}[]
\resizebox{\textwidth}{!}{%
\setlength{\tabcolsep}{5pt}
\begin{tabular}{|c|ccc|ccc|ccc|ccc|}
\hline
\multirow{2}{*}{\textbf{Language}} & \multicolumn{3}{c|}{\multirow{2}{*}{\textbf{Librispeech   English}}}                                                                                        & \multicolumn{3}{c|}{\multirow{2}{*}{\textbf{RESPIN   Bengali}}}                                                                                             & \multicolumn{3}{c|}{\multirow{2}{*}{\textbf{RESPIN   Bhojpuri}}}                                                                                            & \multicolumn{3}{c|}{\multirow{2}{*}{\textbf{Bhashini   Hindi}}}                                                                                             \\
                                   & \multicolumn{3}{c|}{}                                                                                                                                       & \multicolumn{3}{c|}{}                                                                                                                                       & \multicolumn{3}{c|}{}                                                                                                                                       & \multicolumn{3}{c|}{}                                                                                                                                       \\ \hline
\multirow{2}{*}{\textbf{Features}} & \multicolumn{1}{c|}{\multirow{2}{*}{\textbf{LCC ↑}}}        & \multicolumn{1}{c|}{\multirow{2}{*}{\textbf{SRCC ↑}}}       & \multirow{2}{*}{\textbf{MSE ↓}} & \multicolumn{1}{c|}{\multirow{2}{*}{\textbf{LCC ↑}}}        & \multicolumn{1}{c|}{\multirow{2}{*}{\textbf{SRCC ↑}}}       & \multirow{2}{*}{\textbf{MSE ↓}} & \multicolumn{1}{c|}{\multirow{2}{*}{\textbf{LCC ↑}}}        & \multicolumn{1}{c|}{\multirow{2}{*}{\textbf{SRCC ↑}}}       & \multirow{2}{*}{\textbf{MSE ↓}} & \multicolumn{1}{c|}{\multirow{2}{*}{\textbf{LCC ↑}}}        & \multicolumn{1}{c|}{\multirow{2}{*}{\textbf{SRCC ↑}}}       & \multirow{2}{*}{\textbf{MSE ↓}} \\
                                   & \multicolumn{1}{c|}{}                                       & \multicolumn{1}{c|}{}                                       &                                 & \multicolumn{1}{c|}{}                                       & \multicolumn{1}{c|}{}                                       &                                 & \multicolumn{1}{c|}{}                                       & \multicolumn{1}{c|}{}                                       &                                 & \multicolumn{1}{c|}{}                                       & \multicolumn{1}{c|}{}                                       &                                 \\ \hline
\multirow{2}{*}{\textbf{wav2vec 2.0}} & \multicolumn{1}{c|}{\multirow{2}{*}{$87.24 \pm 1.89$}}      & \multicolumn{1}{c|}{\multirow{2}{*}{$89.28 \pm 1.79$}}      & \multirow{2}{*}{$0.016 \pm 0.0027$} & \multicolumn{1}{c|}{\multirow{2}{*}{$80.07 \pm 1.97$}}      & \multicolumn{1}{c|}{\multirow{2}{*}{$83.33 \pm 2.17$}}      & \multirow{2}{*}{$0.021 \pm 0.0020$} & \multicolumn{1}{c|}{\multirow{2}{*}{$79.14 \pm 1.29$}}      & \multicolumn{1}{c|}{\multirow{2}{*}{$82.30 \pm 1.29$}}      & \multirow{2}{*}{$0.023 \pm 0.0018$} & \multicolumn{1}{c|}{\multirow{2}{*}{$80.49 \pm 1.72$}}      & \multicolumn{1}{c|}{\multirow{2}{*}{$80.07 \pm 2.31$}}      & \multirow{2}{*}{$0.018 \pm 0.0013$} \\
                                   & \multicolumn{1}{c|}{}                                       & \multicolumn{1}{c|}{}                                       &                                 & \multicolumn{1}{c|}{}                                       & \multicolumn{1}{c|}{}                                       &                                 & \multicolumn{1}{c|}{}                                       & \multicolumn{1}{c|}{}                                       &                                 & \multicolumn{1}{c|}{}                                       & \multicolumn{1}{c|}{}                                       &                                 \\ \hline
\multirow{2}{*}{\textbf{HuBERT}}   & \multicolumn{1}{c|}{\multirow{2}{*}{$87.33 \pm 1.23$}}      & \multicolumn{1}{c|}{\multirow{2}{*}{$88.97 \pm 1.11$}}      & \multirow{2}{*}{$0.018 \pm 0.0025$} & \multicolumn{1}{c|}{\multirow{2}{*}{$80.56 \pm 1.16$}}      & \multicolumn{1}{c|}{\multirow{2}{*}{$82.90 \pm 1.28$}}      & \multirow{2}{*}{$0.022 \pm 0.0021$} & \multicolumn{1}{c|}{\multirow{2}{*}{$79.50 \pm 1.37$}}      & \multicolumn{1}{c|}{\multirow{2}{*}{$81.54 \pm 1.65$}}      & \multirow{2}{*}{$0.024 \pm 0.0020$} & \multicolumn{1}{c|}{\multirow{2}{*}{$84.01 \pm 1.49$}}      & \multicolumn{1}{c|}{\multirow{2}{*}{$82.62 \pm 1.78$}}      & \multirow{2}{*}{$0.016 \pm 0.0020$} \\
                                   & \multicolumn{1}{c|}{}                                       & \multicolumn{1}{c|}{}                                       &                                 & \multicolumn{1}{c|}{}                                       & \multicolumn{1}{c|}{}                                       &                                 & \multicolumn{1}{c|}{}                                       & \multicolumn{1}{c|}{}                                       &                                 & \multicolumn{1}{c|}{}                                       & \multicolumn{1}{c|}{}                                       &                                 \\ \hline
\multirow{2}{*}{\textbf{PS-I}}     & \multicolumn{1}{c|}{\multirow{2}{*}{$84.61 \pm 2.72$}}      & \multicolumn{1}{c|}{\multirow{2}{*}{$85.42 \pm 3.24$}}      & \multirow{2}{*}{$0.021 \pm 0.0045$} & \multicolumn{1}{c|}{\multirow{2}{*}{$76.18 \pm 3.42$}}      & \multicolumn{1}{c|}{\multirow{2}{*}{$77.64 \pm 3.91$}}      & \multirow{2}{*}{$0.027 \pm 0.0037$} & \multicolumn{1}{c|}{\multirow{2}{*}{$75.89 \pm 3.16$}}      & \multicolumn{1}{c|}{\multirow{2}{*}{$77.14 \pm 4.08$}}      & \multirow{2}{*}{$0.030 \pm 0.0037$} & \multicolumn{1}{c|}{\multirow{2}{*}{$84.17 \pm 1.10$}}      & \multicolumn{1}{c|}{\multirow{2}{*}{$81.74 \pm 0.98$}}      & \multirow{2}{*}{$0.014 \pm 0.0010$} \\
                                   & \multicolumn{1}{c|}{}                                       & \multicolumn{1}{c|}{}                                       &                                 & \multicolumn{1}{c|}{}                                       & \multicolumn{1}{c|}{}                                       &                                 & \multicolumn{1}{c|}{}                                       & \multicolumn{1}{c|}{}                                       &                                 & \multicolumn{1}{c|}{}                                       & \multicolumn{1}{c|}{}                                       &                                 \\ \hline
\multirow{2}{*}{\textbf{PS-II}}    & \multicolumn{1}{c|}{\multirow{2}{*}{$84.72 \pm 1.95$}}      & \multicolumn{1}{c|}{\multirow{2}{*}{$85.44 \pm 2.45$}}      & \multirow{2}{*}{$0.021 \pm 0.0039$} & \multicolumn{1}{c|}{\multirow{2}{*}{$80.06 \pm 2.77$}}      & \multicolumn{1}{c|}{\multirow{2}{*}{$82.22 \pm 3.60$}}      & \multirow{2}{*}{$0.021 \pm 0.0021$} & \multicolumn{1}{c|}{\multirow{2}{*}{$78.59 \pm 2.80$}}      & \multicolumn{1}{c|}{\multirow{2}{*}{$80.10 \pm 3.82$}}      & \multirow{2}{*}{$0.025 \pm 0.0025$} & \multicolumn{1}{c|}{\multirow{2}{*}{$84.98 \pm 2.35$}}      & \multicolumn{1}{c|}{\multirow{2}{*}{$83.83 \pm 2.12$}}      & \multirow{2}{*}{$0.014 \pm 0.0018$} \\
                                   & \multicolumn{1}{c|}{}                                       & \multicolumn{1}{c|}{}                                       &                                 & \multicolumn{1}{c|}{}                                       & \multicolumn{1}{c|}{}                                       &                                 & \multicolumn{1}{c|}{}                                       & \multicolumn{1}{c|}{}                                       &                                 & \multicolumn{1}{c|}{}                                       & \multicolumn{1}{c|}{}                                       &                                 \\ \hline
\multirow{2}{*}{\textbf{PS-III}}   & \multicolumn{1}{c|}{\multirow{2}{*}{$\mathbf{90.93 \pm 1.24}$}} & \multicolumn{1}{c|}{\multirow{2}{*}{$\mathbf{91.62 \pm 1.60}$}} & \multirow{2}{*}{$0.013 \pm 0.0033$} & \multicolumn{1}{c|}{\multirow{2}{*}{$\mathbf{86.64 \pm 2.49}$}} & \multicolumn{1}{c|}{\multirow{2}{*}{$\mathbf{88.69 \pm 2.99}$}} & \multirow{2}{*}{$0.015 \pm 0.0043$} & \multicolumn{1}{c|}{\multirow{2}{*}{$\mathbf{85.49 \pm 3.17}$}} & \multicolumn{1}{c|}{\multirow{2}{*}{$\mathbf{86.88 \pm 3.93}$}} & \multirow{2}{*}{$0.018 \pm 0.0062$} & \multicolumn{1}{c|}{\multirow{2}{*}{$\mathbf{87.21 \pm 0.97}$}} & \multicolumn{1}{c|}{\multirow{2}{*}{$\mathbf{84.93 \pm 0.88}$}} & \multirow{2}{*}{$0.013 \pm 0.0013$} \\
                                   & \multicolumn{1}{c|}{}                                       & \multicolumn{1}{c|}{}                                       &                                 & \multicolumn{1}{c|}{}                                       & \multicolumn{1}{c|}{}                                       &                                 & \multicolumn{1}{c|}{}                                       & \multicolumn{1}{c|}{}                                       &                                 & \multicolumn{1}{c|}{}                                       & \multicolumn{1}{c|}{}                                       &                                 \\ \hline
\end{tabular}
}
\vspace{0.1em} 
\caption{Performance of Proposed Bottleneck Transformer Model on unseen (English, Hindi, Bengali and Bhojpuri) test sets}
\label{tab:unseen1}
\end{table*}

\begin{table*}[!h]
\resizebox{\textwidth}{!}{%
\setlength{\tabcolsep}{5pt}
\begin{tabular}{|c|ccc|ccc|ccc|ccc|}
\hline
\multirow{2}{*}{\textbf{Language}} & \multicolumn{3}{c|}{\multirow{2}{*}{\textbf{Librispeech   English}}}                                                                                    & \multicolumn{3}{c|}{\multirow{2}{*}{\textbf{RESPIN   Bengali}}}                                                                                         & \multicolumn{3}{c|}{\multirow{2}{*}{\textbf{RESPIN   Bhojpuri}}}                                                                                        & \multicolumn{3}{c|}{\multirow{2}{*}{\textbf{Bhashini   Hindi}}}                                                                                         \\
                                   & \multicolumn{3}{c|}{}                                                                                                                                   & \multicolumn{3}{c|}{}                                                                                                                                   & \multicolumn{3}{c|}{}                                                                                                                                   & \multicolumn{3}{c|}{}                                                                                                                                   \\ \hline
\multirow{2}{*}{\textbf{Features}} & \multicolumn{1}{c|}{\multirow{2}{*}{\textbf{LCC ↑}}} & \multicolumn{1}{c|}{\multirow{2}{*}{\textbf{SRCC ↑}}} & \multirow{2}{*}{\textbf{MSE ↓}}          & \multicolumn{1}{c|}{\multirow{2}{*}{\textbf{LCC ↑}}} & \multicolumn{1}{c|}{\multirow{2}{*}{\textbf{SRCC ↑}}} & \multirow{2}{*}{\textbf{MSE ↓}}          & \multicolumn{1}{c|}{\multirow{2}{*}{\textbf{LCC ↑}}} & \multicolumn{1}{c|}{\multirow{2}{*}{\textbf{SRCC ↑}}} & \multirow{2}{*}{\textbf{MSE ↓}}          & \multicolumn{1}{c|}{\multirow{2}{*}{\textbf{LCC ↑}}} & \multicolumn{1}{c|}{\multirow{2}{*}{\textbf{SRCC ↑}}} & \multirow{2}{*}{\textbf{MSE ↓}}          \\
                                   & \multicolumn{1}{c|}{}                                & \multicolumn{1}{c|}{}                                 &                                          & \multicolumn{1}{c|}{}                                & \multicolumn{1}{c|}{}                                 &                                          & \multicolumn{1}{c|}{}                                & \multicolumn{1}{c|}{}                                 &                                          & \multicolumn{1}{c|}{}                                & \multicolumn{1}{c|}{}                                 &                                          \\ \hline
\multirow{2}{*}{\textbf{wav2vec 2.0}} & \multicolumn{1}{c|}{\multirow{2}{*}{$86.22 \pm 3.29$}}   & \multicolumn{1}{c|}{\multirow{2}{*}{$88.20 \pm 3.58$}}    & \multirow{2}{*}{$0.020 \pm 0.0079$}          & \multicolumn{1}{c|}{\multirow{2}{*}{$79.66 \pm 4.20$}}   & \multicolumn{1}{c|}{\multirow{2}{*}{$81.55 \pm 4.47$}}    & \multirow{2}{*}{$0.022 \pm 0.0066$}          & \multicolumn{1}{c|}{\multirow{2}{*}{$77.66 \pm 4.86$}}   & \multicolumn{1}{c|}{\multirow{2}{*}{$79.57 \pm 4.80$}}    & \multirow{2}{*}{$0.026 \pm 0.0071$}          & \multicolumn{1}{c|}{\multirow{2}{*}{$83.26 \pm 1.93$}}   & \multicolumn{1}{c|}{\multirow{2}{*}{$81.60 \pm 1.96$}}    & \multirow{2}{*}{$0.015 \pm 0.0016$}          \\
                                   & \multicolumn{1}{c|}{}                                & \multicolumn{1}{c|}{}                                 &                                          & \multicolumn{1}{c|}{}                                & \multicolumn{1}{c|}{}                                 &                                          & \multicolumn{1}{c|}{}                                & \multicolumn{1}{c|}{}                                 &                                          & \multicolumn{1}{c|}{}                                & \multicolumn{1}{c|}{}                                 &                                          \\ \hline
\multirow{2}{*}{\textbf{HuBERT}}   & \multicolumn{1}{c|}{\multirow{2}{*}{$86.54 \pm 0.86$}}   & \multicolumn{1}{c|}{\multirow{2}{*}{$90.18 \pm 0.81$}}    & \multirow{2}{*}{$0.017 \pm 0.0015$}          & \multicolumn{1}{c|}{\multirow{2}{*}{$81.62 \pm 2.25$}}   & \multicolumn{1}{c|}{\multirow{2}{*}{$85.33 \pm 2.12$}}    & \multirow{2}{*}{$0.019 \pm 0.0022$}          & \multicolumn{1}{c|}{\multirow{2}{*}{$78.19 \pm 1.94$}}   & \multicolumn{1}{c|}{\multirow{2}{*}{$82.84 \pm 1.19$}}    & \multirow{2}{*}{$0.022 \pm 0.0013$}          & \multicolumn{1}{c|}{\multirow{2}{*}{$78.19 \pm 1.94$}}   & \multicolumn{1}{c|}{\multirow{2}{*}{$82.84 \pm 1.19$}}    & \multirow{2}{*}{$0.022 \pm 0.0013$}          \\
                                   & \multicolumn{1}{c|}{}                                & \multicolumn{1}{c|}{}                                 &                                          & \multicolumn{1}{c|}{}                                & \multicolumn{1}{c|}{}                                 &                                          & \multicolumn{1}{c|}{}                                & \multicolumn{1}{c|}{}                                 &                                          & \multicolumn{1}{c|}{}                                & \multicolumn{1}{c|}{}                                 &                                          \\ \hline
\multirow{2}{*}{\textbf{PS-II}}    & \multicolumn{1}{c|}{\multirow{2}{*}{$89.38 \pm 0.68$}}   & \multicolumn{1}{c|}{\multirow{2}{*}{$90.34 \pm 0.50$}}    & \multirow{2}{*}{$\mathbf{0.013 \pm 0.0011}$} & \multicolumn{1}{c|}{\multirow{2}{*}{$85.60 \pm 0.50$}}   & \multicolumn{1}{c|}{\multirow{2}{*}{$88.20 \pm 0.50$}}    & \multirow{2}{*}{$\mathbf{0.015 \pm 0.0005}$} & \multicolumn{1}{c|}{\multirow{2}{*}{$83.70 \pm 0.80$}}   & \multicolumn{1}{c|}{\multirow{2}{*}{$85.60 \pm 0.70$}}    & \multirow{2}{*}{$\mathbf{0.018 \pm 0.0012}$} & \multicolumn{1}{c|}{\multirow{2}{*}{$86.50 \pm 1.40$}}   & \multicolumn{1}{c|}{\multirow{2}{*}{$84.40 \pm 1.10$}}    & \multirow{2}{*}{$\mathbf{0.013 \pm 0.0012}$} \\
                                   & \multicolumn{1}{c|}{}                                & \multicolumn{1}{c|}{}                                 &                                          & \multicolumn{1}{c|}{}                                & \multicolumn{1}{c|}{}                                 &                                          & \multicolumn{1}{c|}{}                                & \multicolumn{1}{c|}{}                                 &                                          & \multicolumn{1}{c|}{}                                & \multicolumn{1}{c|}{}                                 &                                          \\ \hline
\multirow{2}{*}{\textbf{PS-III}}   & \multicolumn{1}{c|}{\multirow{2}{*}{$68.73 \pm 12.53$}}  & \multicolumn{1}{c|}{\multirow{2}{*}{$68.32 \pm 12.11$}}   & \multirow{2}{*}{$0.034 \pm 0.0171$}          & \multicolumn{1}{c|}{\multirow{2}{*}{$77.80 \pm 0.10$}}   & \multicolumn{1}{c|}{\multirow{2}{*}{$79.50 \pm 0.00$}}    & \multirow{2}{*}{$0.024 \pm 0.0032$}          & \multicolumn{1}{c|}{\multirow{2}{*}{$61.90 \pm 8.60$}}   & \multicolumn{1}{c|}{\multirow{2}{*}{$61.70 \pm 10.70$}}   & \multirow{2}{*}{$0.036 \pm 0.0080$}          & \multicolumn{1}{c|}{\multirow{2}{*}{$80.30 \pm 4.10$}}   & \multicolumn{1}{c|}{\multirow{2}{*}{$81.00 \pm 2.30$}}    & \multirow{2}{*}{$0.024 \pm 0.0049$}          \\
                                   & \multicolumn{1}{c|}{}                                & \multicolumn{1}{c|}{}                                 &                                          & \multicolumn{1}{c|}{}                                & \multicolumn{1}{c|}{}                                 &                                          & \multicolumn{1}{c|}{}                                & \multicolumn{1}{c|}{}                                 &                                          & \multicolumn{1}{c|}{}                                & \multicolumn{1}{c|}{}                                 &                                          \\ \hline
\end{tabular}
}
\vspace{0.1em} 
\caption{Performance of the baseline STOI-Net Model on unseen (English, Hindi, Bengali and Bhojpuri) test sets}
\label{tab:unseen2}
\end{table*}

\begin{table}[]
\centering
\setlength{\tabcolsep}{4pt}
\begin{tabular}{|c|c|c|c|c|}
\hline
\textbf{Features}         & \textbf{\#Noise}   & \textbf{LCC↑}                    & \textbf{SRCC↑}                  & \textbf{MSE↓}                      \\ \hline
\multirow{6}{*}{\textbf{wav2vec 2.0}} & \multirow{2}{*}{1} & \multirow{2}{*}{$97.27 \pm 0.55$}  & \multirow{2}{*}{$95.53 \pm 0.60$} & \multirow{2}{*}{$0.0025 \pm 0.0005$} \\
                          &                    &                                  &                                 &                                    \\ \cline{2-5} 
                          & \multirow{2}{*}{2} & \multirow{2}{*}{$94.83 \pm 0.54$}  & \multirow{2}{*}{$95.58 \pm 0.66$} & \multirow{2}{*}{$0.0053 \pm 0.0006$} \\
                          &                    &                                  &                                 &                                    \\ \cline{2-5} 
                          & \multirow{2}{*}{3} & \multirow{2}{*}{$86.47 \pm 0.44$}  & \multirow{2}{*}{$87.91 \pm 0.49$} & \multirow{2}{*}{$0.0104 \pm 0.0003$} \\
                          &                    &                                  &                                 &                                    \\ \hline
\multirow{6}{*}{\textbf{HuBERT}}   & \multirow{2}{*}{1} & \multirow{2}{*}{$97.33 \pm 0.37$}  & \multirow{2}{*}{$93.95 \pm 0.38$} & \multirow{2}{*}{$0.0027 \pm 0.0005$} \\
                          &                    &                                  &                                 &                                    \\ \cline{2-5} 
                          & \multirow{2}{*}{2} & \multirow{2}{*}{$94.55 \pm 0.42$}  & \multirow{2}{*}{$94.86 \pm 0.55$} & \multirow{2}{*}{$0.0056 \pm 0.0005$} \\
                          &                    &                                  &                                 &                                    \\ \cline{2-5} 
                          & \multirow{2}{*}{3} & \multirow{2}{*}{$87.76 \pm 1.66$}  & \multirow{2}{*}{$89.35 \pm 1.43$} & \multirow{2}{*}{$0.0094 \pm 0.0013$} \\
                          &                    &                                  &                                 &                                    \\ \hline
\multirow{6}{*}{\textbf{PS-II}}    & \multirow{2}{*}{1} & \multirow{2}{*}{$95.29 \pm 2.16$}  & \multirow{2}{*}{$90.37 \pm 1.98$} & \multirow{2}{*}{$0.0047 \pm 0.0017$} \\
                          &                    &                                  &                                 &                                    \\ \cline{2-5} 
                          & \multirow{2}{*}{2} & \multirow{2}{*}{$90.15 \pm 1.60$}  & \multirow{2}{*}{$91.82 \pm 1.46$} & \multirow{2}{*}{$0.0097 \pm 0.0016$} \\
                          &                    &                                  &                                 &                                    \\ \cline{2-5} 
                          & \multirow{2}{*}{3} & \multirow{2}{*}{$79.21 \pm 1.48$}  & \multirow{2}{*}{$82.77 \pm 1.58$} & \multirow{2}{*}{$0.0160 \pm 0.0015$} \\
                          &                    &                                  &                                 &                                    \\ \hline
\multirow{6}{*}{\textbf{PS-III}}   & \multirow{2}{*}{1} & \multirow{2}{*}{$97.21 \pm 0.28$}  & \multirow{2}{*}{$93.33 \pm 0.55$} & \multirow{2}{*}{$0.0030 \pm 0.0005$} \\
                          &                    &                                  &                                 &                                    \\ \cline{2-5} 
                          & \multirow{2}{*}{2} & \multirow{2}{*}{$93.64 \pm 0.45$}  & \multirow{2}{*}{$94.19 \pm 0.39$} & \multirow{2}{*}{$0.0066 \pm 0.0005$} \\
                          &                    &                                  &                                 &                                    \\ \cline{2-5} 
                          & \multirow{2}{*}{3} & \multirow{2}{*}{$67.80 \pm 37.31$} & \multirow{2}{*}{$87.51 \pm 0.99$} & \multirow{2}{*}{$0.0117 \pm 0.0006$} \\
                          &                    &                                  &                                 &                                    \\ \hline
\end{tabular}
\vspace{0.1em} 
\caption{Impact on LCC, SRCC and MSE when different number of noises were added in seen test set using proposed model}
\label{tab:noise_count_mse_lcc}
\end{table}

For the proposed model, we used a Conv block for feature extraction, which took input of size 257 for PS-I, 512 for PS-II and PS-III, 768 for wav2vec 2.0 and 1024 for HuBERT features. The hidden dimensions for the first and second Conv1D blocks were 256 and 128, respectively. Both used a kernel size of 3, followed by 1-D Batch Normalization and GELU activation layers. Then the output of the Conv block was given to a bottleneck transformer with a hidden dimension of 64, which took the 128-dimensional input. The bottleneck transformer consisted of 3 blocks, the first block consists of a 2-D convolution layer that had an input dimension of 128, a hidden dimension of 64 and a kernel size of 1 followed by a GELU activation (approximating tanh), 2-D Batch Normalization, and dropout of $0.1$. 
The second block featured an MHSA mechanism that took a 64-dimensional input with 8 heads and a 0.2 dropout. This was followed by a 2-D adaptive average pooling layer, reducing height and width dimensions to $1\times1$ while maintaining the batch size and the number of channels. This was followed by a GELU activation (approximating tanh), 2-D batch normalization, and a 0.1 dropout. The third block included a 2-D convolution layer with input dimensions matching the hidden dimension and output dimensions corresponding to the bottleneck transformer's input, with a kernel size of 1, followed by batch normalization. Finally, a residual connection from the bottleneck input was added, which was passed to the sigmoid activation.
Following the bottleneck transformer, we had Dense Block-1, which included a linear layer with a 128-D input and a 32-D output, followed by layer normalization. We then applied a 1-D adaptive average pooling to eliminate the time dimension. Next, Dense Block-2 is used, featuring a linear layer with a 32-D input and a 1-dimensional output, followed by a sigmoid activation function for the STOI prediction task.

We conducted experiments using the proposed method in PyTorch and the baseline methods in TensorFlow. We used an epoch size of 50, with a learning rate of 0.0001 and Adam as the optimizer. The model parameters and hyper parameters were arrived at experimentally. We used the MSE as the objective function. We used three evaluation metrics: LCC and Spearman rank correlation coefficient
(SRCC)\cite{Klerk2010CommentaryS} to evaluate the performance of the models. MSE was also calculated to compare with the baseline model similar to the work in \cite{zezario2020stoi}. The predictions and comparison of STOI scores were done at an utterance level. 

Furthermore, the performance of the model was analyzed by varying the SNR of the signals, for which the STOI score was predicted. The SNR values of the noisy signals in the Seen test set were calculated using the corresponding clean signals from the dataset as a reference. The SNR values were then grouped into six bins: $ < $ 0dB, 0–5dB, 5–10dB, 10–15dB, 15–20dB and $ > $ 20dB. The plot and table for unseen test data are not shown due to space constraints.

\subsection{Results and Discussion}
In this study, our findings revealed a higher correlation between the ground truth STOI and the predicted STOI scores using the proposed model over the baseline. The average and standard deviation across the 5 folds are shown in \autoref{tab:seen}, \autoref{tab:unseen1}, \autoref{tab:unseen2}, \autoref{tab:noise_count_mse_lcc} and \autoref{tab:snr_mse_lcc}.  

\autoref{tab:param_count} shows the count of trainable parameters across every feature used. It can be seen that the proposed model consistently had fewer parameters than the baseline model across all features.

\autoref{tab:seen} shows the performance on the Seen test set with the cells in bold indicating the highest performance for a metric. The proposed model consistently outperformed the baseline on the seen set. The proposed model achieved the highest LCC $(94.63\pm0.52)$ and SRCC $(95.88\pm0.29)$ values, along with the lowest MSE $(0.0059\pm0.0007)$.

\autoref{tab:unseen1} shows the performance of the proposed architecture on the Unseen test set with different types of input features and \autoref{tab:unseen2} shows that of the baseline. The cells in bold shows the higher performing model for a given metric. \autoref{tab:unseen1} and \autoref{tab:unseen2} show that the proposed model performed better than the baseline, even on unseen data, which highlights its robustness and generalizability. 

The performance on PS-I features are slightly low; it is probably because the model became too simple (~0.33 M parameters) to capture the underlying information present in the spectrogram. By incorporating SSL features, the model demonstrates enhanced performance compared to the baseline. The proposed model shows similar results to the baseline with the PS-II features as with PS-I. The highest performance across the various test sets is observed with the PS-III features for the proposed model. The baseline model showed the highest performance with PS-II as is to be expected.
It can be seen that the proposed performs better with PS-III features compared to the baseline with PS-II features. This gain also comes with a reduction in the number of parameters from $~1.19$M to $~0.67$M. Across all languages and feature types, the proposed architecture gives average LCC, SRCC and MSE of $82.890\pm2.013$, $83.833\pm2.349$ and $0.0195\pm0.0027$, respectively, compared to the baseline counterparts $80.328\pm3.123$, $81.948\pm3.001$ and $0.0212\pm0.0042$, respectively. 

\begin{table}[!t]
\setlength{\tabcolsep}{4pt}
\begin{tabular}{|c|c|c|c|c|}
\hline
\textbf{SNR}                      & \textbf{Features}         & \textbf{LCC↑}                    & \textbf{SRCC↑}                   & \textbf{MSE↓}                      \\ \hline
\multirow{8}{*}{\textbf{\textless{}0}}     & \multirow{2}{*}{Wav2Vec2} & \multirow{2}{*}{$90.83 \pm 1.75$}  & \multirow{2}{*}{$90.63 \pm 1.88$}  & \multirow{2}{*}{$0.0090 \pm 0.0002$} \\
                                  &                           &                                  &                                  &                                    \\ \cline{2-5} 
                                  & \multirow{2}{*}{HuBERT}   & \multirow{2}{*}{$90.95 \pm 1.07$}  & \multirow{2}{*}{$90.87 \pm 1.06$}  & \multirow{2}{*}{$0.0090 \pm 0.0008$} \\
                                  &                           &                                  &                                  &                                    \\ \cline{2-5} 
                                  & \multirow{2}{*}{PS-II}    & \multirow{2}{*}{$83.83 \pm 1.85$}  & \multirow{2}{*}{$83.57 \pm 1.78$}  & \multirow{2}{*}{$0.0166 \pm 0.0014$} \\
                                  &                           &                                  &                                  &                                    \\ \cline{2-5} 
                                  & \multirow{2}{*}{PS-III}   & \multirow{2}{*}{$88.28 \pm 0.41$}  & \multirow{2}{*}{$88.39 \pm 0.39$}  & \multirow{2}{*}{$0.0116 \pm 0.0004$} \\
                                  &                           &                                  &                                  &                                    \\ \hline
\multirow{8}{*}{\textbf{0-5}}              & \multirow{2}{*}{Wav2Vec2} & \multirow{2}{*}{$74.51 \pm 4.46$}  & \multirow{2}{*}{$81.01 \pm 2.71$}  & \multirow{2}{*}{$0.0044 \pm 0.0012$} \\
                                  &                           &                                  &                                  &                                    \\ \cline{2-5} 
                                  & \multirow{2}{*}{HuBERT}   & \multirow{2}{*}{$78.42 \pm 3.65$}  & \multirow{2}{*}{$76.79 \pm 1.78$}  & \multirow{2}{*}{$0.0043 \pm 0.0009$} \\
                                  &                           &                                  &                                  &                                    \\ \cline{2-5} 
                                  & \multirow{2}{*}{PS-II}    & \multirow{2}{*}{$71.45 \pm 8.22$}  & \multirow{2}{*}{$70.98 \pm 4.67$}  & \multirow{2}{*}{$0.0047 \pm 0.0015$} \\
                                  &                           &                                  &                                  &                                    \\ \cline{2-5} 
                                  & \multirow{2}{*}{PS-III}   & \multirow{2}{*}{$72.50 \pm 3.19$}  & \multirow{2}{*}{$73.49 \pm 2.45$}  & \multirow{2}{*}{$0.0043 \pm 0.0011$} \\
                                  &                           &                                  &                                  &                                    \\ \hline
\multirow{8}{*}{\textbf{5-10}}             & \multirow{2}{*}{Wav2Vec2} & \multirow{2}{*}{$76.24 \pm 7.04$}  & \multirow{2}{*}{$79.88 \pm 2.56$}  & \multirow{2}{*}{$0.0047 \pm 0.0015$} \\
                                  &                           &                                  &                                  &                                    \\ \cline{2-5} 
                                  & \multirow{2}{*}{HuBERT}   & \multirow{2}{*}{$72.52 \pm 11.14$} & \multirow{2}{*}{$78.06 \pm 5.67$}  & \multirow{2}{*}{$0.0051 \pm 0.0019$} \\
                                  &                           &                                  &                                  &                                    \\ \cline{2-5} 
                                  & \multirow{2}{*}{PS-II}    & \multirow{2}{*}{$68.20 \pm 6.15$}  & \multirow{2}{*}{$66.63 \pm 6.59$}  & \multirow{2}{*}{$0.0046 \pm 0.0015$} \\
                                  &                           &                                  &                                  &                                    \\ \cline{2-5} 
                                  & \multirow{2}{*}{PS-III}   & \multirow{2}{*}{$73.87 \pm 1.72$}  & \multirow{2}{*}{$76.17 \pm 0.64$}  & \multirow{2}{*}{$0.0043 \pm 0.0007$} \\
                                  &                           &                                  &                                  &                                    \\ \hline
\multirow{8}{*}{\textbf{10-15}}            & \multirow{2}{*}{Wav2Vec2} & \multirow{2}{*}{$62.86 \pm 16.07$} & \multirow{2}{*}{$72.07 \pm 14.61$} & \multirow{2}{*}{$0.0013 \pm 0.0010$} \\
                                  &                           &                                  &                                  &                                    \\ \cline{2-5} 
                                  & \multirow{2}{*}{HuBERT}   & \multirow{2}{*}{$61.22 \pm 14.91$} & \multirow{2}{*}{$57.47 \pm 14.29$} & \multirow{2}{*}{$0.0012 \pm 0.0002$} \\
                                  &                           &                                  &                                  &                                    \\ \cline{2-5} 
                                  & \multirow{2}{*}{PS-II}    & \multirow{2}{*}{$49.34 \pm 16.74$} & \multirow{2}{*}{$43.38 \pm 10.44$} & \multirow{2}{*}{$0.0029 \pm 0.0025$} \\
                                  &                           &                                  &                                  &                                    \\ \cline{2-5} 
                                  & \multirow{2}{*}{PS-III}   & \multirow{2}{*}{$65.57 \pm 4.33$}  & \multirow{2}{*}{$63.00 \pm 4.64$}  & \multirow{2}{*}{$0.0012 \pm 0.0003$} \\
                                  &                           &                                  &                                  &                                    \\ \hline
\multirow{8}{*}{\textbf{15-20}}            & \multirow{2}{*}{Wav2Vec2} & \multirow{2}{*}{$26.51 \pm 1.00$}  & \multirow{2}{*}{$50.40 \pm 4.46$}  & \multirow{2}{*}{$0.0063 \pm 0.0018$} \\
                                  &                           &                                  &                                  &                                    \\ \cline{2-5} 
                                  & \multirow{2}{*}{HuBERT}   & \multirow{2}{*}{$48.14 \pm 7.32$}  & \multirow{2}{*}{$50.13 \pm 5.07$}  & \multirow{2}{*}{$0.0018 \pm 0.0004$} \\
                                  &                           &                                  &                                  &                                    \\ \cline{2-5} 
                                  & \multirow{2}{*}{PS-II}    & \multirow{2}{*}{$29.64 \pm 8.66$}  & \multirow{2}{*}{$16.73 \pm 8.13$}  & \multirow{2}{*}{$0.0035 \pm 0.0026$} \\
                                  &                           &                                  &                                  &                                    \\ \cline{2-5} 
                                  & \multirow{2}{*}{PS-III}   & \multirow{2}{*}{$38.09 \pm 6.83$}  & \multirow{2}{*}{$32.16 \pm 5.03$}  & \multirow{2}{*}{$0.0019 \pm 0.0004$} \\
                                  &                           &                                  &                                  &                                    \\ \hline
\multirow{8}{*}{\textbf{\textgreater{}20}} & \multirow{2}{*}{Wav2Vec2} & \multirow{2}{*}{$46.23 \pm 18.54$} & \multirow{2}{*}{$37.15 \pm 11.69$} & \multirow{2}{*}{$0.0011 \pm 0.0006$} \\
                                  &                           &                                  &                                  &                                    \\ \cline{2-5} 
                                  & \multirow{2}{*}{HuBERT}   & \multirow{2}{*}{$43.60 \pm 14.78$} & \multirow{2}{*}{$41.63 \pm 8.91$}  & \multirow{2}{*}{$0.0021 \pm 0.0003$} \\
                                  &                           &                                  &                                  &                                    \\ \cline{2-5} 
                                  & \multirow{2}{*}{PS-II}    & \multirow{2}{*}{$5.43 \pm 10.23$}  & \multirow{2}{*}{$25.76 \pm 9.35$}  & \multirow{2}{*}{$0.0056 \pm 0.0027$} \\
                                  &                           &                                  &                                  &                                    \\ \cline{2-5} 
                                  & \multirow{2}{*}{PS-III}   & \multirow{2}{*}{$16.99 \pm 10.09$} & \multirow{2}{*}{$31.26 \pm 6.85$}  & \multirow{2}{*}{$0.0031 \pm 0.0016$} \\
                                  &                           &                                  &                                  &                                    \\ \hline
\end{tabular}
\vspace{0.4em} 
\caption{Impact on LCC, SRCC, and MSE across different SNR levels in the seen test set using proposed model}
\label{tab:snr_mse_lcc}
\end{table}

A detailed noise-specific result (increasing levels of noise contamination) is presented in \autoref{tab:noise_count_mse_lcc} for different feature extraction methods. The observations show that with increasing the number of noises, the correlation of the actual and predicted STOI scores reduce while the MSE values increase. This shows an intuitive drift in the prediction performance because intelligibility of speech reduces as more noises are added to a clean recording. 

In order to answer the question of how STOI predictions vary when signal quality varies based on SNR, the noisy Indic TIMIT data was binned into various SNR bins. Correlation scores between the predicted and actual STOI scores were calculated for each bin for the various features used in the experiment. It was observed that the correlation score of signals having a lower SNR ($<$10dB) were higher than the correlation scores of signals with higher SNR ($>$20dB). Initially this seemed counter to the results in \autoref{tab:snr_mse_lcc}; however, the plots in \autoref{fig:predicted_vs_actual_stoi_across_snr_bins_W2V2_Features} show how the predicted STOI and actual STOI scores vary for signals in different SNR ranges. As can be seen, for lower SNR ranges, there is a spread in the actual and predicted values that is approximately linear, resulting in a high correlation. As signal quality improves (for higher SNR ranges), the prediction and actual STOI scores tend to concentrate in a smaller region. This results in a lack of any linear trend in their relationship, resulting in a poor correlation. This further shows that there is a complex relation of intelligibility with the number of noises in a signal and the SNR of the respective signal.

\section{Conclusions}
In this study, we proposed a model architecture to predict the STOI scores, which is a non-intrusive speech intelligibility assessment model and does not require the clean reference signal for STOI prediction. We have leveraged the bottleneck transformer as the backbone for our model to learn the local and global context present in the data. We have used SSL features (wav2vec 2.0, HuBERT) and spectral features for our experiments. Experimental results indicate that our proposed model consistently achieved higher LCC and SRCC values across both seen and unseen test sets, demonstrating improved accuracy, reliability, and generalizability. We also observed that for lower SNR ($<$10dB), actual and predicted STOI scores are more highly correlated than for higher SNR. In the future, we plan to use adapter-based fine-tuning of SSL features to further improve STOI predictions. There is also scope for exploring predictions using more recent models like Whisper or Conformer-based models with the option of having the predictions that include a combination of metrics.

\section*{Acknowledgement}

The authors express their sincere gratitude to the Centre for Artificial Intelligence and Robotics (CAIR), Defence Research and Development Organisation (DRDO), Government of India, for their support and encouragement throughout this work.

\bibliographystyle{IEEEtran}
\bibliography{mybib}

@inproceedings{fu2018quality,
  title={{Quality-Net: An end-to-end non-intrusive speech quality assessment model based on blstm}},
  author={{Fu, Szu-Wei and Tsao, Yu and Hwang, Hsin-Te and Wang, Hsin-Min}},
  booktitle={Proc. Interspeech},
  year={2018}}

@INPROCEEDINGS{941023,
  author={{Rix, A.W. and Beerends, J.G. and Hollier, M.P. and Hekstra, A.P.}},
  booktitle={Proc. ICASSP}, 
  title={{Perceptual evaluation of speech quality (PESQ) -a new method for speech quality assessment of telephone networks and codecs}}, 
  year={2001},
  volume={2},
  number={},
  pages={749-752 vol.2},
  doi={10.1109/ICASSP.2001.941023}
}

@inproceedings{zezario2020stoi,
  title={{STOI-Net: A deep learning based non-intrusive speech intelligibility assessment model}},
  author={{Zezario, Ryandhimas E and Fu, Szu-Wei and Fuh, Chiou-Shann and Tsao, Yu and Wang, Hsin-Min}},
  booktitle={Asia-Pacific Signal and Information Processing Association Annual Summit and Conference (APSIPA ASC)},
  pages={482--486},
  year={2020},
  organization={IEEE}
}

@inproceedings{zezario122022mti,
  title     = {{MTI-Net: A Multi-Target Speech Intelligibility Prediction Model}},
  author    = {{Ryandhimas Edo Zezario and Szu-wei Fu and Fei Chen and Chiou-Shann Fuh and Hsin-Min Wang and Yu Tsao}},
  year      = {{2022}},
  journal = {},
  booktitle = {{Interspeech 2022}},
  pages     = {{5463--5467}},
  doi       = {{10.21437/Interspeech.2022-10828}},
  issn      = {{2958-1796}},
}

@article{zezario2022deep,
  title={{Deep learning-based non-intrusive multi-objective speech assessment model with cross-domain features}},
  author={{Zezario, Ryandhimas E and Fu, Szu-Wei and Chen, Fei and Fuh, Chiou-Shann and Wang, Hsin-Min and Tsao, Yu}},
  journal={IEEE/ACM Transactions on Audio, Speech, and Language Processing},
  volume={31},
  pages={54--70},
  year={2022},
  publisher={IEEE}
}

@inproceedings{taal2010short,
  title={{A short-time objective intelligibility measure for time-frequency weighted noisy speech}},
  author={{Taal, Cees H and Hendriks, Richard C and Heusdens, Richard and Jensen, Jesper}},
  booktitle={Proc. ICASSP},
  pages={4214--4217},
  year={2010},
  organization={IEEE}
}

@article{hsu2021hubert,
  title={{Hubert: Self-supervised speech representation learning by masked prediction of hidden units}},
  author={{Hsu, Wei-Ning and Bolte, Benjamin and Tsai, Yao-Hung Hubert and Lakhotia, Kushal and Salakhutdinov, Ruslan and Mohamed, Abdelrahman}},
  journal={IEEE/ACM transactions on audio, speech, and language processing},
  volume={29},
  pages={3451--3460},
  year={2021},
  publisher={IEEE}
}

@article{baevski2020wav2vec,
  title={{wav2vec 2.0: A framework for self-supervised learning of speech representations}},
  author={{Baevski, Alexei and Zhou, Yuhao and Mohamed, Abdelrahman and Auli, Michael}},
  journal={Advances in neural information processing systems},
  volume={33},
  pages={12449--12460},
  year={2020}
}

@article{kates2010hearing,
  title={{The hearing-aid speech quality index (HASQI)}},
  author={{Kates, James M and Arehart, Kathryn H}},
  journal={Journal of the Audio Engineering Society},
  volume={58},
  number={5},
  pages={363--381},
  year={2010},
  publisher={Audio Engineering Society}
}

@article{kates2014hearing,
  title={{The Hearing-Aid Speech Perception Index (HASPI)}},
  author={{Kates, James M and Arehart, Kathryn H}},
  journal={Speech Communication},
  volume={65},
  pages={75--93},
  year={2014},
  publisher={Elsevier BV}
}

@inproceedings{edozezario22_interspeech,
  title     = {{MBI-Net: A Non-Intrusive Multi-Branched Speech Intelligibility Prediction Model for Hearing Aids}},
  author    = {{Ryandhimas {Edo Zezario} and Fei Chen and Chiou-Shann Fuh and Hsin-Min Wang and Yu Tsao}},
  year      = {2022},
  booktitle = {Proc. Interspeech},
  pages     = {3944--3948},
  doi       = {10.21437/Interspeech.2022-10838},
  issn      = {2958-1796},
}

@inproceedings{zezario24_interspeech,
  title     = {{Non-Intrusive Speech Intelligibility Prediction for Hearing Aids using Whisper and Metadata}},
  author    = {{Ryandhimas E. Zezario and Fei Chen and Chiou-Shann Fuh and Hsin-Min Wang and Yu Tsao}},
  year      = {2024},
  booktitle = {Proc. Interspeech},
  pages     = {3844--3848},
  doi       = {10.21437/Interspeech.2024-716},
  issn      = {2958-1796},
}

@inproceedings{lin2024non,
  title={{A Non-Intrusive Speech Quality Assessment Model using Whisper and Multi-Head Attention}},
  author={{Lin, Guojian and Tsao, Yu and Chen, Fei}},
  booktitle={2024 Asia Pacific Signal and Information Processing Association Annual Summit and Conference (APSIPA ASC)},
  pages={1--6},
  year={2024},
  organization={IEEE}
}

@INPROCEEDINGS{9577771,
  author={{Srinivas, Aravind and Lin, Tsung-Yi and Parmar, Niki and Shlens, Jonathon and Abbeel, Pieter and Vaswani, Ashish}},
  booktitle={2021 IEEE/CVF Conference on Computer Vision and Pattern Recognition (CVPR)}, 
  title={{Bottleneck Transformers for Visual Recognition}}, 
  year={2021},
  volume={},
  number={},
  pages={16514-16524},
}

@INPROCEEDINGS{yarra2019indic,
  author={{Yarra, Chiranjeevi and Aggarwal, Ritu and Rajpal, Avni and Ghosh, Prasanta Kumar}},
  booktitle={Proc. O-COCOSDA}, 
  title={{Indic TIMIT and Indic English lexicon: A speech database of Indian speakers using TIMIT stimuli and a lexicon from their mispronunciations}}, 
  year={2019},
  volume={},
  number={},
  pages={1-6},
  doi={10.1109/O-COCOSDA46868.2019.9041230}
}

@INPROCEEDINGS{panayotov2015librispeech,
  author={{Panayotov, Vassil and Chen, Guoguo and Povey, Daniel and Khudanpur, Sanjeev}},
  booktitle={Proc. ICASSP}, 
  title={{Librispeech: An ASR corpus based on public domain audio books}}, 
  year={2015},
  volume={},
  number={},
  pages={5206-5210},
  doi={10.1109/ICASSP.2015.7178964}
}

@misc{hendrycks2023gaussianerrorlinearunits,
      title={{Gaussian Error Linear Units (GELUs)}}, 
      author={{Dan Hendrycks and Kevin Gimpel}},
      year={2023},
      eprint={1606.08415},
      archivePrefix={arXiv},
      primaryClass={cs.LG},
      url={https://arxiv.org/abs/1606.08415}, 
}

@misc{musan2015,
  author = {{David Snyder and Guoguo Chen and Daniel Povey}},
  title = {{{MUSAN}: {A} {M}usic, {S}peech, and {N}oise {C}orpus}},
  year = {2015},
  eprint = {1510.08484},
  note = {arXiv:1510.08484v1}
}

@inproceedings{kim08e_interspeech,
  title     = {{Robust signal-to-noise ratio estimation based on waveform amplitude distribution analysis}},
  author    = {{Chanwoo Kim and Richard M. Stern}},
  year      = {2008},
  booktitle = {Proc. Interspeech},
  pages     = {2598--2601},
  doi       = {10.21437/Interspeech.2008-644},
  issn      = {2958-1796},
}

@article{Klerk2010CommentaryS,
  title={{Commentary: Spearman’s ‘The proof and measurement of association between two things’}},
  author={{Nicholas de Klerk}},
  journal={International Journal of Epidemiology},
  year={2010},
  volume={39},
  pages={1159-1161},
  url={https://api.semanticscholar.org/CorpusID:196423054}
}

@INPROCEEDINGS{10447597,
  author={{Mogridge, Rhiannon and Close, George and Sutherland, Robert and Hain, Thomas and Barker, Jon and Goetze, Stefan and Ragni, Anton}},
  booktitle={Proc. ICASSP}, 
  title={{Non-Intrusive Speech Intelligibility Prediction for Hearing-Impaired Users Using Intermediate ASR Features and Human Memory Models}}, 
  year={2024},
  volume={},
  number={},
  pages={306-310},
  doi={10.1109/ICASSP48485.2024.10447597}}

@article{DRGAS2024103068,
title = {{Speech intelligibility prediction using generalized ESTOI with fine-tuned parameters}},
journal = {Speech Communication},
volume = {159},
pages = {103068},
year = {2024},
issn = {0167-6393},
doi = {https://doi.org/10.1016/j.specom.2024.103068},
url = {https://www.sciencedirect.com/science/article/pii/S0167639324000402},
author = {{Szymon Drgas}},
}

@ARTICLE{10870355,
  author={{Zhou, Xiajie and Mawalim, Candy Olivia and Unoki, Masashi}},
  journal={IEEE Access}, 
  title={{Speech Intelligibility Prediction Using Binaural Processing for Hearing Loss}}, 
  year={2025},
  volume={},
  number={},
  pages={1-1},
  doi={10.1109/ACCESS.2025.3538708}}

@ARTICLE{7486010,
  author={{Eaton, James and Gaubitch, Nikolay D. and Moore, Alastair H. and Naylor, Patrick A.}},
  journal={IEEE/ACM Transactions on Audio, Speech, and Language Processing}, 
  title={{Estimation of Room Acoustic Parameters: The ACE Challenge}}, 
  year={2016},
  volume={24},
  number={10},
  pages={1681-1693},
  doi={10.1109/TASLP.2016.2577502}}

@inproceedings{tamm22_interspeech,
  title     = {{Pre-trained Speech Representations as Feature Extractors for Speech Quality Assessment in Online Conferencing Applications}},
  author    = {{Bastiaan Tamm and Helena Balabin and Rik Vandenberghe and Hugo {Van hamme}}},
  year      = {2022},
  booktitle = {Proc. Interspeech},
  pages     = {4083--4087},
  doi       = {10.21437/Interspeech.2022-10147},
  issn      = {2958-1796},
}

@INPROCEEDINGS{9746395,
  author={{Cooper, Erica and Huang, Wen-Chin and Toda, Tomoki and Yamagishi, Junichi}},
  booktitle={Proc. ICASSP}, 
  title={{Generalization Ability of MOS Prediction Networks}}, 
  year={2022},
  volume={},
  number={},
  pages={8442-8446},
  doi={10.1109/ICASSP43922.2022.9746395}}

@INPROCEEDINGS{leng_2021_mbnet,
  author={{Leng, Yichong and Tan, Xu and Zhao, Sheng and Soong, Frank and Li, Xiang-Yang and Qin, Tao}},
  booktitle={Proc. ICASSP}, 
  title={{MBNET: MOS Prediction for Synthesized Speech with Mean-Bias Network}}, 
  year={2021},
  volume={},
  number={},
  pages={391-395},
  doi={10.1109/ICASSP39728.2021.9413877}}

@inproceedings{lo2019mosnet,
  title     = {{MOSNet: Deep Learning-Based Objective Assessment for Voice Conversion}},
  author    = {{Chen-Chou Lo and Szu-Wei Fu and Wen-Chin Huang and Xin Wang and Junichi Yamagishi and Yu Tsao and Hsin-Min Wang}},
  year      = {2019},
  booktitle = {Proc. Interspeech},
  pages     = {1541--1545},
  doi       = {10.21437/Interspeech.2019-2003},
  issn      = {2958-1796},
}

@INPROCEEDINGS{singh2023model,
  author={{Udupa, Sathvik and Bandekar, Jesuraja and Deekshitha, G and Kumar, Saurabh and Ghosh, Prasanta Kumar and Badiger, Sandhya and Singh, Abhayjeet and Murthy, Savitha and Pai, Priyanka and Raghavan, Srinivasa and Nanavati, Raoul}},
  booktitle={IEEE Automatic Speech Recognition and Understanding Workshop (ASRU)}, 
  title={{Gated Multi Encoders and Multitask Objectives for Dialectal Speech Recognition in Indian Languages}}, 
  year={2023},
  volume={},
  number={},
  pages={1-8},
  doi={10.1109/ASRU57964.2023.10389624}}

@ARTICLE{wavLM_paper,
  author={{Chen, Sanyuan and Wang, Chengyi and Chen, Zhengyang and Wu, Yu and Liu, Shujie and Chen, Zhuo and Li, Jinyu and Kanda, Naoyuki and Yoshioka, Takuya and Xiao, Xiong and Wu, Jian and Zhou, Long and Ren, Shuo and Qian, Yanmin and Qian, Yao and Wu, Jian and Zeng, Michael and Yu, Xiangzhan and Wei, Furu}},
  journal={IEEE Journal of Selected Topics in Signal Processing}, 
  title={{WavLM: Large-Scale Self-Supervised Pre-Training for Full Stack Speech Processing}}, 
  year={2022},
  volume={16},
  number={6},
  pages={1505-1518},
  doi={10.1109/JSTSP.2022.3188113}}

@inproceedings{kondo2025rethinking,
  title={Rethinking Mean Opinion Scores in Speech Quality Assessment: Score Aggregation through Quantized Distribution Fitting},
  author={Kondo, Yuto and Kameoka, Hirokazu and Tanaka, Kou and Kaneko, Takuhiro},
  booktitle={Proc. ICASSP},
  pages={1--5},
  year={2025},
  organization={IEEE}
}

@ARTICLE{clipping_9281027,
  author={Záviška, Pavel and Rajmic, Pavel and Ozerov, Alexey and Rencker, Lucas},
  journal={IEEE Journal of Selected Topics in Signal Processing}, 
  title={A Survey and an Extensive Evaluation of Popular Audio Declipping Methods}, 
  year={2021},
  volume={15},
  number={1},
  pages={5-24},
  doi={10.1109/JSTSP.2020.3042071}}

\vspace{12pt}
\color{red}

\end{document}